\documentclass{ifacconf}

\usepackage{graphicx}      
\usepackage{natbib}        

\usepackage{epstopdf}
\usepackage{url}
\usepackage{multirow}

\begin{document}
\begin{frontmatter}

\title{PID2018 Benchmark Challenge: Model-based Feedforward Compensator with A Conditional Integrator \thanksref{footnoteinfo}}

\thanks[footnoteinfo]{This work was supported by China Scholarship Council (CSC) under Grant(201606090086).}
\thanks[footnoteinfo]{Corresponding author: Jie Yuan (jieyuan@seu.edu.cn) or YangQuan Chen (ychen53@ucmerced.edu).}

\author[First,Fifth]{Jie Yuan}
\author[Second]{Abdullah Ates}
\author[Third]{Sina Dehghan}
\author[Fourth]{Yang Zhao}
\author[Fifth]{Shumin Fei}
\author[Third]{YangQuan Chen}

\address[First]{School of Automation, Southeast University, Nanjing
210096, China (e-mail: {\tt{jieyuan@seu.edu.cn}}).}
\address[Second]{Engineering Faculty, Computer Engineering Department,Inonu University, Malatya, 44280, Turkey (e-mail:{\tt{ abdullah.ates@inonu.edu.tr}}).}
\address[Third]{Mechatronics, Embedded Systems and Automation
Lab, University of California, Merced, CA 95340, USA (e-mail: {\tt{sdehghan@ucmerced.edu}}; {\tt{ychen53@ucmerced.edu}})}
\address[Fourth]{School of Science and Engineering, Shandong
University, Jinan 250061, China, (e-mail: {\tt{zdh1136@gmail.com}})}
\address[Fifth]{Key Laboratory of Measurement and
Control of CSE, Ministry of Education, School of Automation, Southeast
University, Nanjing 210096, China, (e-mail: {\tt{smfei@seu.edu.cn}})}

\begin{abstract}                
Since proportional-integral-derivative (PID) controllers absolutely dominate the control engineering, numbers of different control structures and theories have been developed to enhance the efficiency of PID controllers. Thus, it is essential and inspiring to operate different PID control strategies to the PID2018 Benchmark Challenge. In this paper, a novel control strategy is designed for this refrigeration system, where a feedforward compensator and a conditional integrator are utilized to compensate the disturbances and remove the steady-state error in the benchmark problem, respectively. The simulation results given in the benchmark problem show the straightforward effectiveness of the proposed control structure compared with the existing control methods.
\end{abstract}


\end{frontmatter}

\section{Introduction}
Temperature control plays an important role in the process industry. In cooling generation, vapour compression based refrigeration is now the leading technology worldwide. The energy consumed in the heating and cooling processes accounts for a large part of the total energy consumption. It is reported that about 30\% of total energy over the world contributes to the heating, ventilating, and air conditioning, as well as the refrigerators and water heaters \citep{Jahangeer2011}, where refrigerators occupy 28\% of home energy consumption in the United States \citep{steemers2009household}. As energy efficiency is one of the most powerful weapons for combating global climate change, it is necessary to not only control the refrigeration systems precisely, but also in a more efficient way.

As is known, the refrigeration system is a closed cycle. Its components are connected through various pipes and valves, which causes strong nonlinearities and high coupling. That is why the dynamic modelling of vapour-compression refrigeration systems is not a trivial matter. The heat exchanger is the most important element regarding the dynamic modelling, while the expansion valve, the compressor, and the thermal behaviour of the secondary fluxes can be statically modelled since their dynamics are usually at least one order of magnitude faster than those of the evaporator and condenser.

It is known that feedforward control plays a significant role in disturbance rejection. In practice, some of the disturbances are measurable or pre-known, which can be utilized to compensate the disturbances completely and improve the system performance. In the feedforward control structure, the control signal is not based on the tracking error, but on the mathematical model of the process and the measurement of the disturbance. Feedforward control for disturbance rejection has been widely used in industry, such as disk drives \citep{Jahangeer2011} and high-precision motion control \citep{su2004disturbance}. Though the plant model and the disturbance model are assumed to be exactly accurate, it is not always possible to remove the disturbance perfectly. It is known that a feedforward controller always contains the inverse of the plant model. Therefore, when the plant model has non-minimum phase zeros or the plant model delay is larger than the delay of the disturbance path model, it is not acceptable or achievable to inverse these elements, which will lead to the instability or non-causality of the controller. To solve this problem, \cite{zhong2012feedforward} discussed the stable and causal approximation of the feedforward controller.

Motivated by aforementioned issues, a novel control strategy will be provided for the refrigeration system. On one hand, a feedforward controller will be designed to compensate the disturbances. On the other hand, the conditional integrator will be introduced to reduce the phase lag while maintaining the steady-state accuracy. A detailed simulation example will be provided to show the effectiveness of the proposed strategy.

The remainder of this paper is organized as follows: section 2 describes the controlled plant in view of the control aspect. In section 3, the system modeling work is introduced. The detailed design procedures of the feedforward controller are presented in section 4. The conditional integrator is utilized to reduce the steady-error in section 5. The simulation results shown in section 6 demonstrate the effectiveness of the proposed structure. The paper is concluded in section 7.

\section{Refrigeration system description}
\begin{figure}[htbp]
\begin{center}
\includegraphics[width=8.4cm]{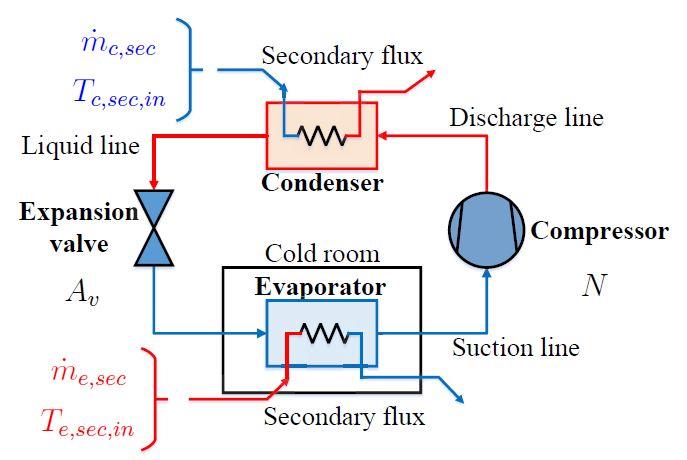}    
\caption{Schematic picture of one-compression-stage, one-load-demand vapour- compression refrigeration cycle \citep{Bejarano2018}.}
\label{plant}
\end{center}
\end{figure}

The canonical one-compression-stage, one-load-demand refrigeration cycle is shown in Fig. \ref{plant} and the system variables are given in Table \ref{variables}.

\begin{table}[htbp]
\centering
\caption{Refrigeration system variables}
\label{variables}
\begin{tabular}{|c|c|c|}
\hline
\multicolumn{2}{|c|}{Variables} & { Range} \\ \hline
\multirow{2}{*}{Manipulated variables}  & $A_v$  & 10-100 \ \% \\ \cline{2-3}
                   & $N$  & 30-50 \ \rm{Hz} \\ \hline
\multirow{2}{*}{Output variables}  &  $T_{sec,evap,out}$ &  - \\ \cline{2-3}
                   & $T_{SH}$  & - \\ \hline
\end{tabular}
\end{table}

The output variable $T_{e,sec,out}$ is the outlet temperature of the evaporator secondary flux. The highest evaporator efficiency will be achieved when the refrigerant at the evaporator outlet is saturated vapour. However, this behaviour is not acceptable in practice, since the temperature of the evaporator outlet is very high in transient. The risk of liquid droplets will appear since the evaporator outlet matches the compressor intake, which must be definitely avoided. Thus, the superheating of the refrigerant at the evaporator outlet $T_{SH}$, is designed to be another reference to be tracked. The manipulated variable $A_v$ is the expansion valve opening position in Fig. \ref{plant}, and $N$ is the compressor speed. Seven variables stated in the PID2018 Benchmark documentation \citep{Bejarano2018} can be considered as disturbances. The expansion valve, the compressor, and the thermal behavior of secondary fluxes are statically modelled since their dynamics are usually much faster than those of the evaporator and condenser. The major disturbances are the inlet temperature of the evaporator secondary flux $T_{e,sec,in}$ and the inlet temperature of the condenser secondary flux $T_{c,sec,in}$. These are the disturbances to be compensated in the control objective. The initial operating point of the manipulated variables, output variables, and these two major disturbances are indicated in Table \ref{initial}.

\begin{table}[htbp]
\centering
\caption{Initial operating point}
\label{initial}
\begin{tabular}{|c|c|c|}
\hline
\multicolumn{2}{|c|}{Variables}                    &           range            \\ \hline
\multirow{2}{*}{Manipulated variables} &         $A_v$              &       $\cong 48.79\ \%$                 \\ \cline{2-3}
                  &          $N$             &        $\cong 36.45\  \rm{Hz}$                \\ \hline
\multirow{2}{*}{Output variables} &            $T_{e,sec,out}$ & $\cong -22.15\ ^{\circ} \rm{C}$ \\ \cline{2-3}
                  & $T_{SH}$                            & $\cong 14.65\ ^{\circ} \rm{C}$ \\ \hline
\multirow{2}{*}{Disturbances} &             $T_{c,sec,in}$          &       $30\  ^{\circ} \rm{C}$              \\ \cline{2-3}
                  &          $T_{e,sec,in}$             &            $-20\ ^{ \circ }\rm{C}$           \\ \hline
\end{tabular}
\end{table}

\section{model identification for the refrigeration system}
 Several modeling works have already been done by \cite{macarthur1983application,mckinley2008advanced,li2010dynamic,pangborn2015comparison} and all of them are in detailed review. However, the structures are very complicated. As this system is a black box in the simulation model, it is too difficult to build an accurate model. To simplify the modeling work, we used the input step change to build the transfer functions of this system in Simulink form. 
  Moreover, based on the different step changes in the system identification, the system model is shown to be nonlinear since it will change with different system inputs. However, it is impossible to build the system model at each different input value. In this paper, only the nominal model is identified. A lookup table is generated which represents the steady-state gain. 
  For the nominal model case, the manipulated variable step change is chosen as, $A_v=58.79\%$ at 200 (sec), (58.79\% lies in the middle of the working range), and $N=41.45\rm{Hz}$ at time 200 (sec) respectively, because the system output is definitely steady at 200 (sec), and the step response will be stabilized before 540 (sec) (when the system disturbances appear).

%

The nominal models from $A_v$ to $T_{sec,e,out}$ and from $N$ to $T_{SH}$ are identified as

\begin{equation}
G_{T_{e,sec,out},A_{v}}=\frac{-0.6325s-0.01147}{s^2+16.87s+0.6216},
\end{equation}

\begin{equation}
G_{T_{SH},N}=\frac{3.662s+0.07604}{s^2+19.63s+0.4441}.
\end{equation}

Due to the nonlinearity of the system, the model dynamics varies when the manipulated variable input changes. For the simplicity, a lookup table in reference with the steady-state gain is generated in order to scale the model gain to the real response steady-state gain. Several points for $A_v$ and $N$ are picked in the working range. Under these different step inputs, one can calculate the response steady-state gain, and times the inverse of the nominal gain. On this basis, the lookup table data can be generated. The break points are the difference between the picked value and the initial value of $A_v$ or $ N$. These two lookup tables for two transfer functions are given as Table \ref{lookup2} and Table \ref{lookup1}.

\begin{table}[hb]
\begin{center}
\caption{Lookup table for $G_{T_{e,sec,out}-A_v}(0)$ }\label{lookup2}
\begin{tabular}{cc}
Break points & Table data \\\hline
-38.79 & 2.0981 \\
-30 & 1.6986 \\
-20 & 1.3550 \\
-10 & 1.2282 \\
10 & 1.0000 \\
20 & 0.9106 \\
30 & 0.8336 \\
40 & 0.7664 \\
51.21 & 0.6997\\ \hline
\end{tabular}
\end{center}
\end{table}

\begin{table}[hb]
\begin{center}
\caption{Lookup table for $G_{T_{SH}-N}(0)$} \label{lookup1}
\begin{tabular}{cc}
Break points & Table data \\\hline
-6.45 & 1.2895 \\
-5 & 1.2423 \\
5 & 1.0000\\
13.55 & 0.8569 \\ \hline
\end{tabular}
\end{center}
\end{table}

\section{Disturbance Feedforward compensation}
\begin{figure}[htbp]
\begin{center}
\includegraphics[width=8.4cm]{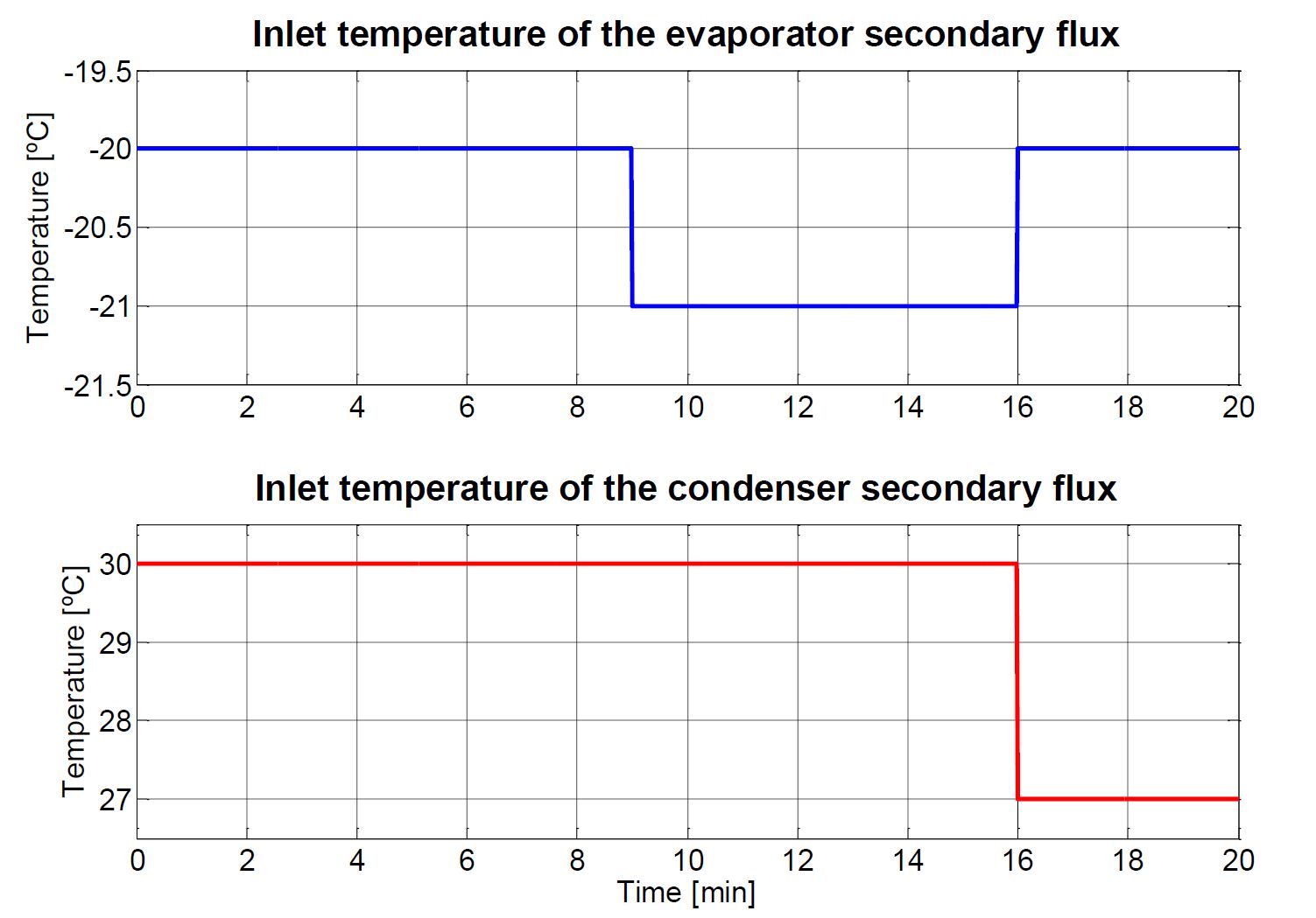}    
\caption{The standard simulation for PID2018 Benchmark generates changes in two
disturbances: $T_{e,sec,in}$ and $T_{c,sec,in}$.}
\label{Disturbances}
\end{center}
\end{figure}

For the refrigerator system, there are seven parameters can be viewed as the disturbances. It is noted that the most important element regarding to the dynamic modeling is the heat exchanger, while the expansion valve, compressor, and the thermal behavior of the secondary fluxes can be statically modeled. Five of these seven parameters are constant numbers throughout the {\rm{SIMULINK}} operation and they are injected parameters in the modeling other than disturbances. The main disturbances are the inlet temperatures of the evaporator secondary flux $T_{e,sec,in}$ and the condenser secondary flux $T_{c,sec,in}$. These two parameters have step changes as shown in Fig. \ref{Disturbances}.

\begin{figure}[htbp]
\begin{center}
\includegraphics[width=8.4cm]{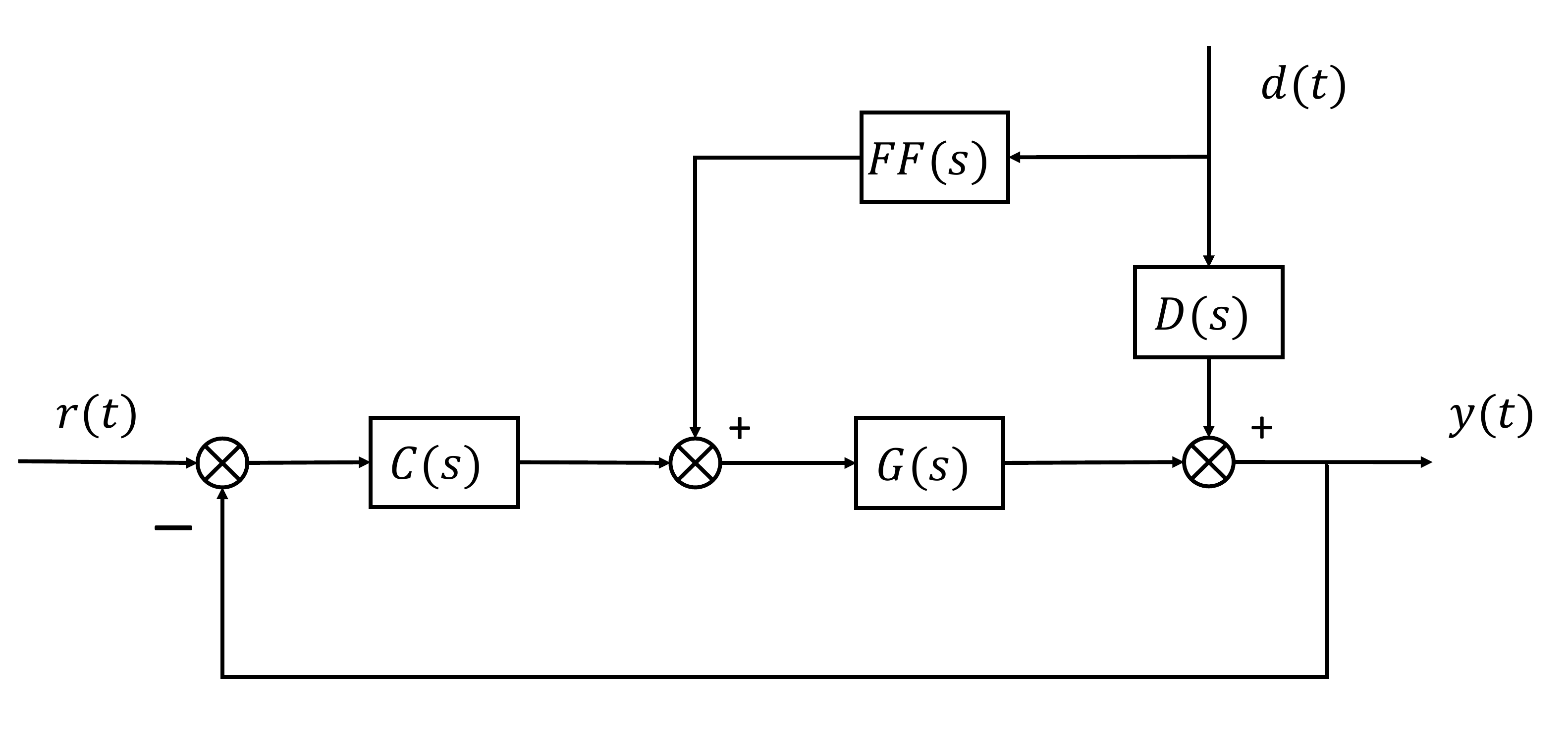}    
\caption{The feedforward control structure of disturbance rejection for SISO system}
\label{FFdiagram}
\end{center}
\end{figure}

Feedforward compensation technique is widely used in disturbance rejection. The disturbance feedforward control diagram for a Single-Input-Single-Output (SISO) system is shown in Fig. \ref{FFdiagram}, where $C(s)$ is the controller, $G(s)$ is the plant, $d(t)$ is the disturbance signal, $D(s)$ is the disturbance path model, $FF(s)$ is the feedforward compensator, $r(t)$ is the system reference, and $y(t)$ is the system output.

The feedforward controller is designed to eliminate the effect of the disturbance signal to the system output. On this basis, it is derived that
\begin{equation}
D_t(s)FF(s)G(s)+D_t(s)D(s)=0,
\end{equation}
where $D_t(s)$ is Laplace transform of the disturbance signal $d(t)$. Thus, the feedforward compensator is derived as
\begin{equation}   \label{FF}
FF(s)=-\frac{D(s)}{G(s)}.
\end{equation}

Although feedforward control is a very mature technique in control theory, there are few references or tutorials that show the detailed procedures of designing a feedforward controller. In the following, the feedforward controller design method will be introduced in detail. Firstly and most importantly, model the disturbance path $D(s)$. There are two disturbances in this Benchmark problem. As shown in Fig. \ref{Disturbances}, $T_{e,sec,in}$ has a step decrease at 540 (sec) and a step increase at 960 (sec), while $T_{c,sec,in}$ has a step decrease at 960 (sec). When modeling the disturbance path, only one disturbance should be implemented to the system, thus another disturbance should be compensated as a constant value at first. In addition, the disturbance path modeling should be done in the open-loop situation, which means the controller does not work in the modeling part. Taking the disturbance rejection $T_{e,sec,in}$ as an example, the feedforward compensator design procedures are as follows

\begin{enumerate}
  \item Cut off the two feedbacks in the closed-loop system, and set the reference signals of $T_{e,sec,out}$ and $T_{SH}$ as zero to make sure the controllers do not work.
  \item Compensate the disturbance $T_{c,sec,in}$ as a constant value, which equals 30.
  \item Implement the disturbance $T_{e,sec,in}$ into the system.
  \item Capture the transient processes of the output $T_{e,sec,out}$ and $T_{SH}$ and model the two step responses to the disturbance step change, which is $-3$ at 540 (sec), respectively.
  \item Generate two feedforward compensators according to Equation (\ref{FF}).
  \item Add the compensation signals to their corresponding controller signals.
 \end{enumerate}

%

Repeating above procedures for disturbance rejection of $T_{c,sec,out}$, four disturbance path models are given as
\begin{equation}
D_{11}(s)= \frac{44.84}{s+45.58},
\end{equation}

\begin{equation}
D_{21}(s)= \frac{-109.1s+4.903}{s^2+256.4s+7.268},
\end{equation}

\begin{equation}
D_{12}(s)= \frac{0.008624}{s+0.04323},
\end{equation}

\begin{equation}
D_{22}(s)= \frac{0.572}{s+0.04099},
\end{equation}
where $D_{11}(s)$ represents the disturbance path from the first disturbance $T_{e,sec,in}$ to the first output $T_{e,sec,out}$, $D_{12}(s)$ represents the disturbance path from the second disturbance $T_{c,sec,in}$ to the first output $T_{e,sec,out}$, $D_{21}(s)$ represents the disturbance path from the first disturbance $T_{e,sec,in}$ to the second output $T_{SH}$, and $D_{22}(s)$ represents the disturbance path from the second disturbance $T_{c,sec,in}$ to the second output $T_{SH}$. Because the controllers are all in discrete time, the Feedforward controllers should also be implemented in discrete time. Hence, the compensators are regenerated as digital filters with sampling period 1 (sec)
\begin{equation}
F_{11}=\frac{0.1268z^2-0.02234z-0.09628}{z^2-1.94z+0.9405},
\end{equation}

\begin{equation}
F_{21}=\frac{-0.1655z^2+0.02693z+0.1319}{z^2-1.939z+0.9401},
\end{equation}

\begin{equation}
F_{12}=\frac{28.32z^2-4.989z-21.5}{z^2-0.0661z-0.8995},
\end{equation}

\begin{equation}
F_{22}=\frac{2.404z^3-2.905z^2-1.506z+2.003}{z^3--0.967z^2-0.9692z+0.9373}.
\end{equation}

Then $F_{11}$ and $F_{21}$ can be added after the first controller to compensate the manipulated signal $A_v$. $F_{12}$ and $F_{22}$ can be added after the second controller to compensate the manipulated signal $N$.

\section{Conditional Integrator}
It is widely known that the integrator is used to remove the steady-state error of the response in control engineering. However, the linear integrator contains $90^\circ$ phase lag at all frequencies, which deteriorates the control performances and may even lead to instability. A conditional integrator known as the Clegg integrator, was proposed by \cite{Clegg1958p4142} to reduce the phase lag while maintaining the steady-state accuracy. Clegg integrator is widely used in reset control systems \citep{Banos2011} and the model of it is given as
\begin{equation}
u(t) = \left\{ {\begin{array}{*{20}{c}}
{\int_{{t_0}}^t {e(v)dv\quad } }&{e(t) \ne 0}\\
0&{e(t) = 0}
\end{array}} \right.,
\end{equation}

where, the integrator output is reset to zero immediately when the error $e(t)$ changes sign. The Clegg integrator acts like a linear integrator whenever its output and input have the same sign. Otherwise, the output is reset to zero.
It is easily noticed that in the PID2018 Benchmark, the system outputs, even for the response baseline, generally have non-zero steady-state errors, which influence the performance in terms of the index $J$. In order to remove the steady-state error, a conditional integrator inspired by the Clegg integrator is introduced as follows

\begin{equation}
u(t) = \left\{ {\begin{array}{*{20}{c}}
{w\int_{{t_0}}^t {{e^*}(\tau )} d\tau }&\qquad {e(t) \ne 0}\\
0&\qquad {e(t) = 0}
\end{array}} \right..
\end{equation}

\begin{figure}[htbp]
\begin{center}
\includegraphics[width=8.4cm]{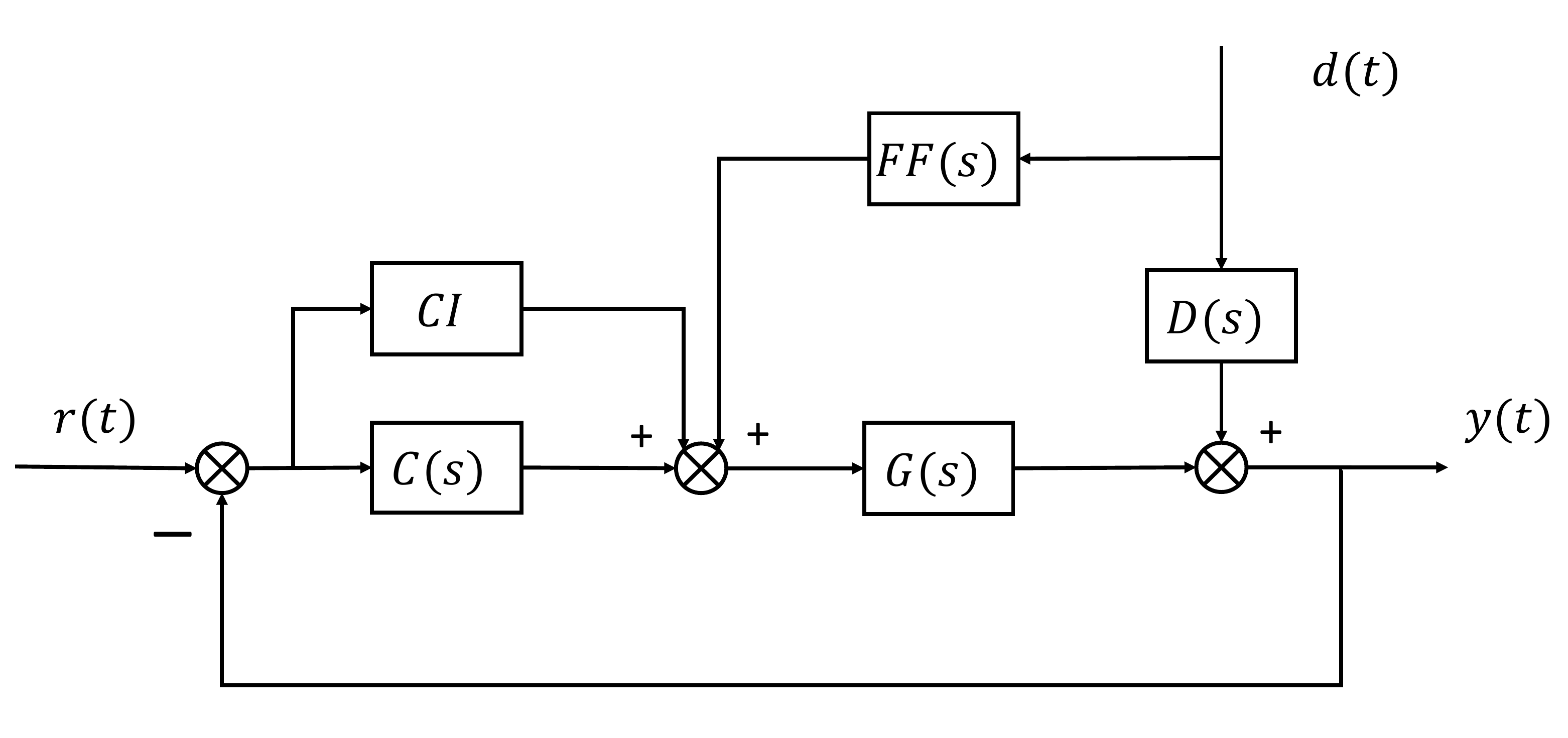}    
\caption{The feedforward control structure with conditional integrator for SISO system}
\label{FF+CI}
\end{center}
\end{figure}

Here,
\begin{equation}
{e^ * }(\tau ) = \left\{ {\begin{array}{*{20}{c}}
{e(\tau )}&&\qquad{\left| {e(\tau )} \right| \le \delta }\\
0&&\qquad{{\rm{otherwise}}}
\end{array}} \right.,
\end{equation}

where, $\delta$ is the threshold of the integrated error, $w$ is the weight parameter. Other Conditional integrator methods can also be used here, such as those in \cite{luo2010tuning}. When the absolute value of the error between the reference signal and output is smaller than the threshold $\delta$, the conditional integrator will generate an additional signal which is added to the control signal to accelerate the convergence speed to the reference. The control structure combined with conditional integrator and feedforward compensator for SISO system is shown in Fig. \ref{FF+CI}. In this paper, the parameters $w$ and $\delta$ are chosen manually and separately for the $T_{e,sec,out}-A_v$ and $T_{SH}-N$ loops.

\section{Simulation results}
Based on the feedforward control strategy introduced previously, three controllers are defined, Controller 1 (C1) is the is the default controllers in \cite{Bejarano2018}, Controller 2 (C2) is the combination of the default controllers and feedforward compensators, Controller 3 (C3) is the is the combination of the default controllers, feedforward compensators, and conditional integrators.

\begin{figure}[htbp]
\begin{center}
\includegraphics[width=9.4cm,height=5cm]{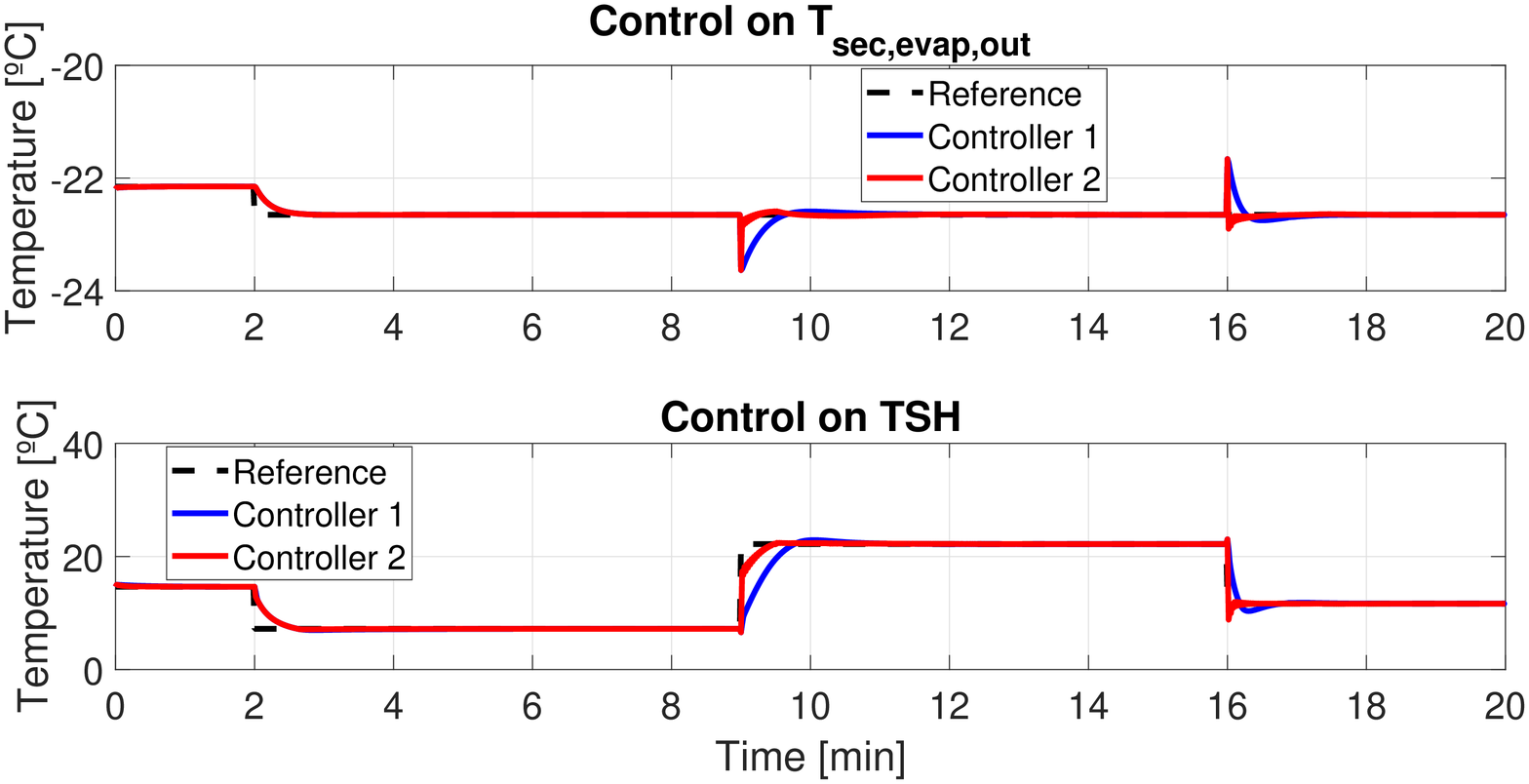}    
\caption{Qualitative comparison of two control structures with the MIMO Refrigeration Control System. Controlled variables.}
\label{yout}
\end{center}
\end{figure}

\begin{figure}[htbp]
\begin{center}
\includegraphics[width=9.4cm]{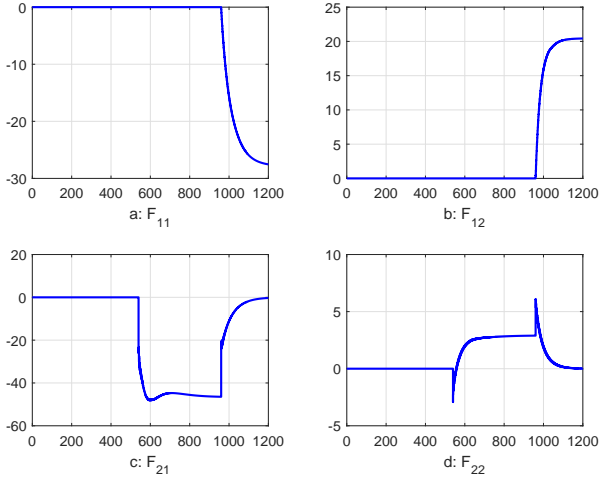}    
\caption{The feedforward compensation signals.}
\label{FFsignal}
\end{center}
\end{figure}

\begin{figure}[htbp]
\begin{center}
\includegraphics[width=9.4cm,height=5cm]{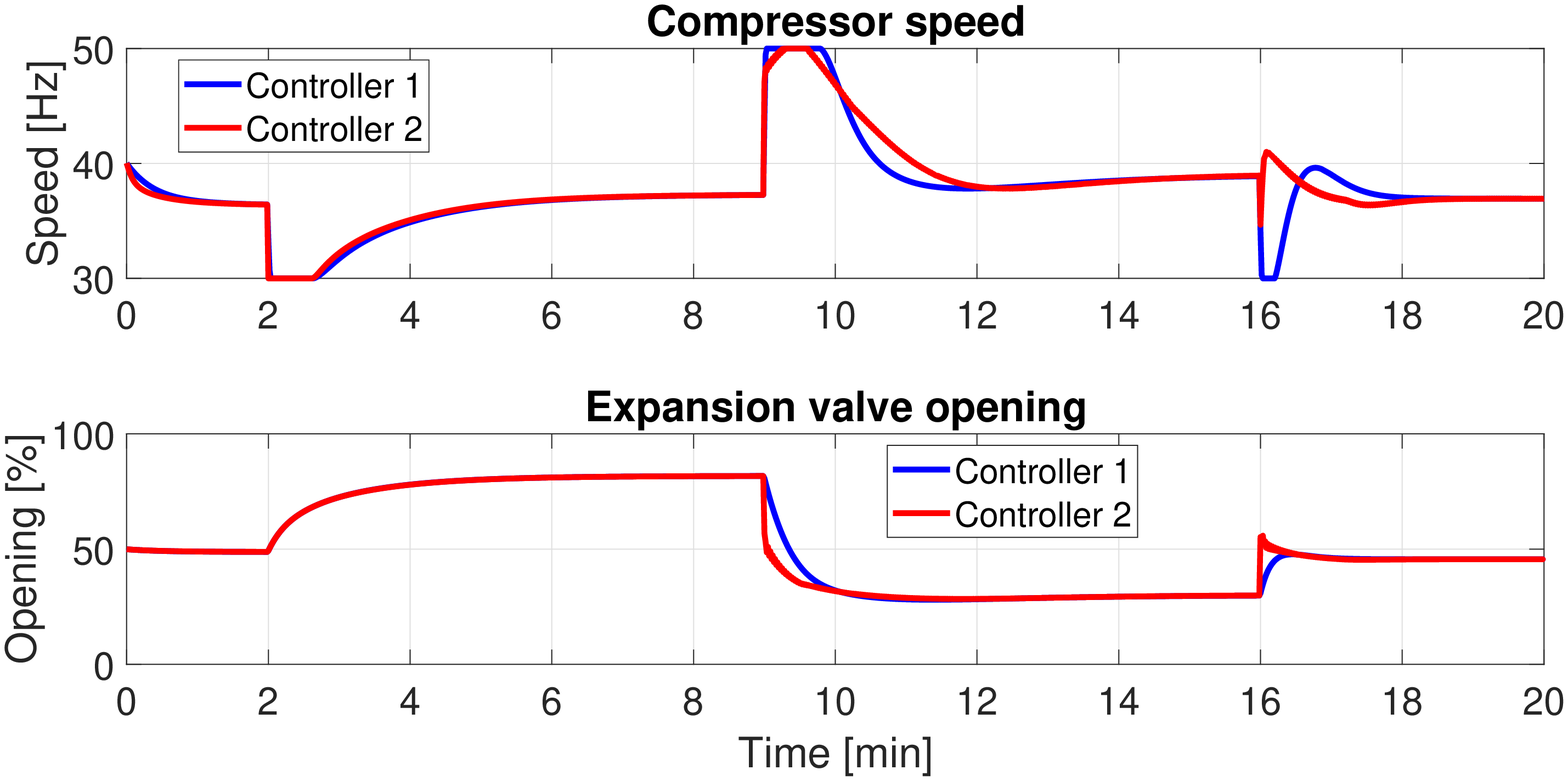}    
\caption{Qualitative comparison of two control structure with the MIMO Refrigeration Control System. Manipulated variables. }
\label{control}
\end{center}
\end{figure}

The system time responses for C1 are shown in Fig. \ref{yout}. The compensation signals generated by the feedforward controllers are shown in Fig. \ref{FFsignal} and the total manipulated signals are shown in Fig. \ref{control}. The quantitative indexes are given in Table \ref{index}. From Fig. \ref{yout}, when the disturbances occur at 540 (sec) and 960 (sec), the responses of C2 departure the reference signal and then quickly return back to the set-point, which is ascribed to the compensation signal that accelerate the disturbance rejection. It can be noticed from Fig. \ref{control} that the total manipulated signals $A_v$ and $N$ have a much shorter saturated duration compared with that of C1 after the feedforward compensation. Although the feedforward controller $F_{12}$ is not injected to the system, the compensation signal is also plotted in Fig. \ref{FFsignal} (b). One can find that it reaches to 20, which will make $N$ get saturated immediately and will last until the end. Hence, we remove this signal from the control structure.

\begin{table}[htb]
\begin{center}
\caption{Relative indices and the combined index associated to the qualitative controller
comparison between C1 and C2.}\label{index}
\begin{tabular}{cc}
Index & Value\\\hline
${\rm{RIAE}}_1({\rm{C_2,C_1}})$ & 0.4482 \\
${\rm{RIAE}}_2({\rm{C_2,C_1}})$ & 0.5188 \\
${\rm{RITAE}}_1({\rm{C_2,C_1,t_{c1},t_{s1}}})$ & 1.0003 \\
${\rm{RITAE}}_2({\rm{C_2,C_1,t_{c2},t_{s2}}})$ & 0.9999 \\
${\rm{RIAE}}_2({\rm{C_2,C_1,t_{c3},t_{s3}}})$ & 0.7236 \\
${\rm{RIAE}}_2({\rm{C_2,C_1,t_{c4},t_{s4}}})$ & 0.3720 \\
${\rm{RIAVU_1}}_1({\rm{C_2,C_1}})$ & 1.7204 \\
${\rm{RIAVU_2}}_1({\rm{C_2,C_1}})$ & 1.1452 \\ \hline
$J({\rm{C_2,C_1}})$ & 0.7445 \\ \hline
\end{tabular}
\end{center}
\end{table}

\begin{figure}
\begin{center}
\includegraphics[width=8.4cm]{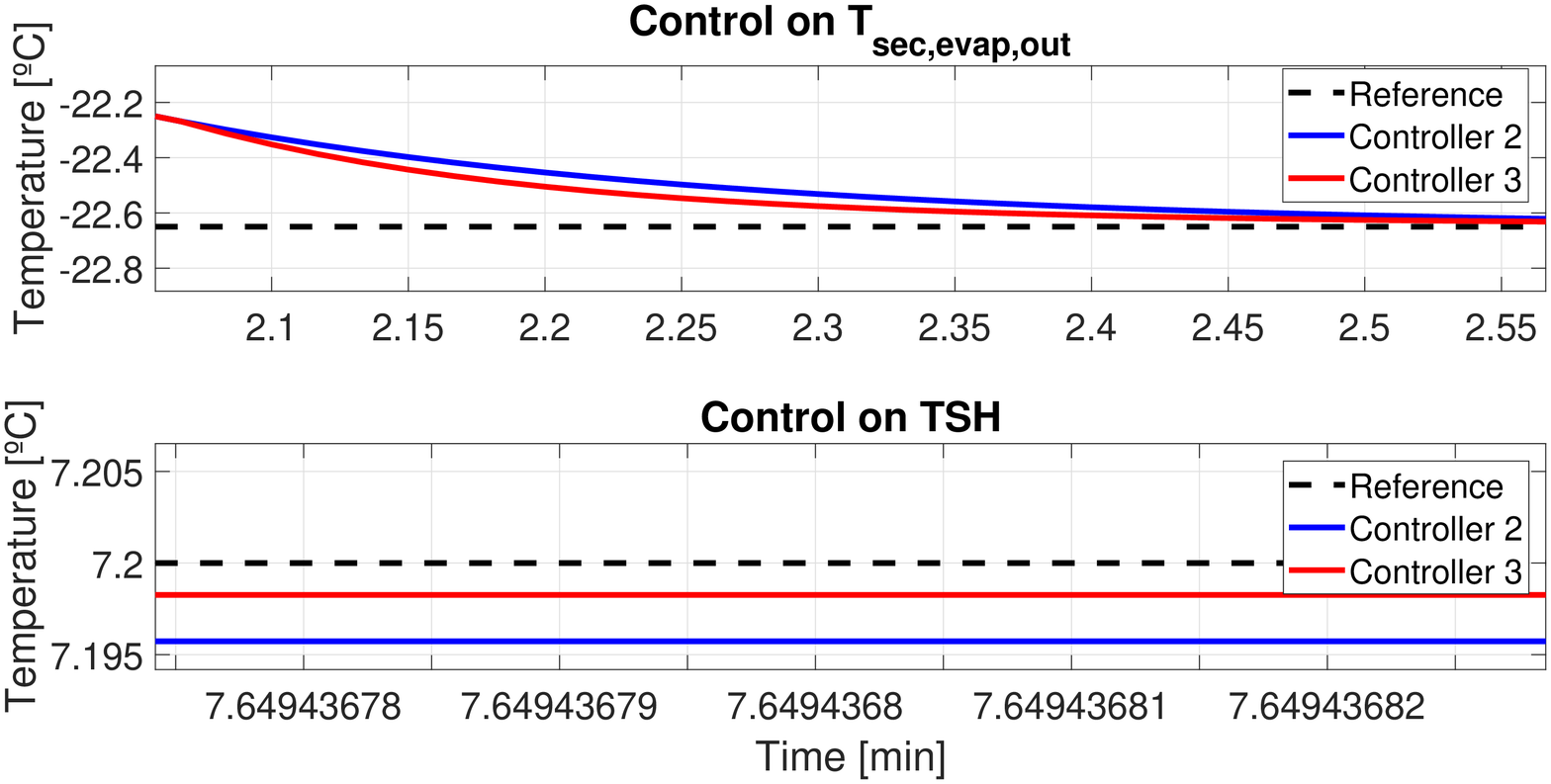}    
\caption{Qualitative comparison between C2 and C3 with the MIMO Refrigeration Control System. Controlled variables.}
\label{yout2}
\end{center}
\end{figure}

Based on the results of the feedforward compensation, the threshold and weight parameters are initially chosen to be 1. Manually decrease these values by using trial and error methodology until find out the relatively better index $J$. The conditional integrator parameters are shown in Table \ref{CIpara}. The system responses for C3 are shown in Fig. \ref{yout2}. The response $T_{e,sec,out}$ shows an obvious convergence when the tracking error equals the threshold $\delta =0.4$. The steady-state error of $T_{SH}$ have also reduced to 0.0023 from 0.0045. The quantitative indexes are given in Table \ref{index2}. The indexes comparison between C1 and C3 are given in Table \ref{index3}. The final comparison index $J$ indicates that the proposed control structure can improve the basic performance by 56.6\%.

\begin{table}[htb]
\centering
\caption{The parameters of the conditional integrators}
\label{CIpara}
\begin{tabular}{|c|c|c|}
\hline
 Control loop& Threshold $\delta$ & Weight parameter $w$ \\ \hline
 $T_{e,sec,out}$-$A_v$  & 0.4 & 1 \\ \hline
 $T_{SH}$-$N$& 0.5 & 0.5 \\ \hline
\end{tabular}
\end{table}

\begin{table}[htb]
\begin{center}
\caption{Relative indices and the combined index associated to the qualitative controller
comparison between C2 and C3.}\label{index2}
\begin{tabular}{cc}
Index & Value\\\hline
${\rm{RIAE}}_1({\rm{C_3,C_2}})$ & 0.9058 \\
${\rm{RIAE}}_2({\rm{C_3,C_2}})$ & 0.7794 \\
${\rm{RITAE}}_1({\rm{C_3,C_2,t_{c1},t_{s1}}})$ & 0.7303 \\
${\rm{RITAE}}_2({\rm{C_3,C_2,t_{c2},t_{s2}}})$ & 0.5574 \\
${\rm{RIAE}}_2({\rm{C_3,C_2,t_{c3},t_{s3}}})$ & 0.5088 \\
${\rm{RIAE}}_2({\rm{C_3,C_2,t_{c4},t_{s4}}})$ & 0.6032 \\
${\rm{RIAVU_1}}_1({\rm{C_3,C_2}})$ & 1.0168 \\
${\rm{RIAVU_2}}_1({\rm{C_3,C_2}})$ & 1.2045 \\ \hline
$J({\rm{C_3,C_2}})$ & 0.7517 \\ \hline
\end{tabular}
\end{center}
\end{table}

\begin{table}[htb]
\begin{center}
\caption{Relative indices and the combined index associated to the qualitative controller
comparison between C1 and C3.}\label{index3}
\begin{tabular}{cc}
Index & Value\\\hline
${\rm{RIAE}}_1({\rm{C_3,C_1}})$ & 0.4060 \\
${\rm{RIAE}}_2({\rm{C_3,C_1}})$ & 0.4043 \\
${\rm{RITAE}}_1({\rm{C_3,C_1,t_{c1},t_{s1}}})$ & 0.7305 \\
${\rm{RITAE}}_2({\rm{C_3,C_1,t_{c2},t_{s2}}})$ & 0.5573 \\
${\rm{RIAE}}_2({\rm{C_3,C_1,t_{c3},t_{s3}}})$ & 0.3682 \\
${\rm{RIAE}}_2({\rm{C_3,C_1,t_{c4},t_{s4}}})$ & 0.2244 \\
${\rm{RIAVU_1}}_1({\rm{C_3,C_1}})$ & 1.7494 \\
${\rm{RIAVU_2}}_1({\rm{C_3,C_1}})$ & 1.7494 \\ \hline
$J({\rm{C_3,C_1}})$ & 0.5662 \\ \hline
\end{tabular}
\end{center}
\end{table}

\section{Conclusion}
This paper focuses on the PID2018 Benchmark program. The canonical one-compression-stage, one-load-demand refrigeration cycle system is identified at first and two second-order transfer functions are derived. Then the disturbances are utilized to set up feedforward controllers to compensate the disturbances and the detailed feedforward controller design procedures are also presented. The conditional integrator is used to accelerate the system output convergence speed and reduce the steady-state error. The simulation results finally show the effectiveness of feedforward compensators in terms of the disturbance rejection, and the benefits of conditional integrator in terms of the steady-state error.

\bibliography{BenchmarkPID}             








\end{document}